\documentclass{ws-ijgmmp}
\usepackage{notoccite}
\usepackage[usenames,dvipsnames,svgnames]{xcolor}
\usepackage[colorlinks=true,
      linkcolor=red,
      urlcolor=gray,
      citecolor=blue]{hyperref}
\usepackage[]{cite}
\usepackage{float}
\usepackage{subfig}
\def \D {\tilde{\nabla}}
\def\ber {\begin{eqnarray}}
\def\eer {\end{eqnarray}}

\begin{document}

\markboth{Heba Sami, Neo Namane, Joseph Ntahompagaze, Maye Elmardi, Amare Abebe}
{Reconstructing $f(R)$ Gravity  from a Chaplygin Scalar Field in de Sitter Spacetimes}

\catchline{}{}{}{}{}
%
%%%%%%%%%%%%%%%%%%%%%%%%%%%%%%%%%%%%%%%%%%%%%%%%%%%%%%%%%%%%%%%%%%%%

\title{Reconstructing $f(R)$ Gravity  from a Chaplygin Scalar Field in de Sitter Spacetimes}

\author{Heba Sami }

\address{Department of Physics, North-West University, Mafikeng, South Africa,\\
Center for Space Research, North-West University, South Africa\\
hebasami.abdulrahman@gmail.com
}

\author{Neo Namane }

\address{Department of Physics, North-West University, Mafikeng, South Africa,\\
Center for Space Research, North-West University, South Africa
}

\author{Joseph Ntahompagaze }

\address{Astronomy and Astrophysics Division, Entoto Observatory and Research Center,\\
Addis Ababa, Ethiopia,\\
Department of Physics, College of Science and Technology, University of Rwanda, \\
 Kigali, Rwanda,\\
 Department of Physics, North-West University, Mafikeng, South Africa}

 \author{Maye Elmardi }

\address{Department of Physics, North-West University, Mafikeng, South Africa,\\
Gravity and Cosmology Group, Department of Mathematics and Applied Mathematics, University of Cape Town, Cape Town, South Africa
}
 
\author{Amare Abebe}

\address{Department of Physics, North-West University, \\
Mafikeng, South Africa\\
Center for Space Research, North-West University, South Africa
}

\maketitle

\begin{abstract}
We present a reconstruction technique for models of $f(R)$ gravity from the Chaplygin scalar field in flat de Sitter spacetimes.
Exploiting the equivalence between $f(R)$ gravity and scalar-tensor theories, and treating the Chaplygin gas as a scalar field model 
in a universe without conventional matter forms, the Lagrangian densities for the $f(R)$ action are derived. Exact $f(R)$ models  
and corresponding scalar field potentials are obtained for asymptotically de Sitter spacetimes in early and late cosmological  
expansion histories. It is shown that the reconstructed $f(R)$ models all have General Relativity as a limiting solution.
\end{abstract}

\ccode{PACS: 04.50.Kd, 04.25.Nx, 98.80.-k, 95.36.+x, 98.80.Cq}

\section{Introduction}
The standard model of cosmology with $\Lambda$CDM fits a number of observational data very well \cite{kow08}. The rapid development of observational cosmology recently has shown that the Universe has undergone two phases of cosmic accelerated expansion. The first one is the so-called inflation \cite{staro80,sato81,kaz80} which is understood to have occurred prior to the radiation-dominated era \cite{lid00}. This phase is required not only to solve the flatness and horizon problems plagued in standard Big Bang cosmology, but also to explain a nearly flat spectrum of temperature anisotropies observed in the CMB \cite{smoot92}. The second accelerating expansion phase has started after the matter-domination era, which implies that the pressure $p$ and the energy density $\mu$ of the universe should have violated the strong physical energy condition $\mu+3p >0$ \cite{ries98}. Therefore, this late cosmic acceleration cannot be explained by the presence of standard matter
whose equation of state $w=p/\mu$ satisfies the condition $w \geq 0$. In fact, we need a component of negative pressure with at least $w < -1/3 $ to realise the acceleration of the universe. This unknown smooth component responsible for this acceleration in the expansion rate is referred to as ``dark energy'' \cite{sahn00}. The need for the existence of dark energy has been confirmed by a number of observations such as SN Ia \cite{ries98,ries99}, the LSS \cite{teg04,teg06}, the BAO \cite{perc07}, and the CMB \cite{Kom09}.

However, there are numerous recent attempts to explain away dark energy. Two of the most common such attempts come either in the form of  modifications to the theory of gravity or the introduction of new matter or scalar field contributions to the action of General Relativity (GR).  In the latter case, one suggestion is that the change in the behavior of the missing energy density might be regulated by the change in the equation of state of the background fluid instead of the form of the potential \cite{chatto10,bento2002generalized}. The Chaplygin Gas (CG) model in cosmology is one of the most profound candidates for this suggestion. For quite sometime now, the CG model has been considered as another alternative to the cosmological FLRW universe models with a perfect fluid equation of state and negative pressure \cite{gorini04,dev03}. The model provides interesting features of the cosmic expansion history consistent with a smooth transition between an inflationary phase and a matter-dominated decelerating era. In addition, the late-time accelerated de Sitter phase of cosmic expansion can be achieved \cite{kamenshchik2001alternative,bilic02,gorini03,ben12}.

Among the earliest and  simplest modifications to the GR gravitational action are $f(R)$ gravity models. These models are usually considered to be geometrical alternatives to the dark energy debate \cite{carroll04,capoz06,capozziello2011extended,nojiri2011unified,nojir11,noj07,sot10,clifton2011modified,elmardi16}, but their scopes of applicability has only been increasing: from early-universe cosmic inflation \cite{staro80,nojiri03,nojiri07}, to the evolutionary dynamics of large-scale structure \cite{song07,abebe13,dombriz08,carloni08,ananda09,elmardi15,abebe16qN,abebe14aN,abebe16ani} and astrophysics \cite{capozziello2008extended,dombrizphd,chiba07,alvaro16}.

An interesting aspect of $f(R)$ gravitational models is their proven equivalence to a class of the Brans-Dicke (BD) version of scalar-tensor (ST) theories \cite{sot06,fara07,faulk07,jos17}. The main objective of this paper is to make use of this equivalence and reconstruct models of $f(R)$ gravity that describe exactly the same background cosmological evolution as Chaplygin gas models that mimic a BD scalar field with a vanishing coupling constant. 

The remaining part of the manuscript is organised as follows: in Section \ref{fRChapST}, we will give a brief description 
of the three underlying alternative cosmological theories used for our analysis, namely, $f(R)$ models of gravity, 
the Chaplygin gas as a cosmological solution and the BD classes of ST theories. After solving the background fluid equations 
for the Chaplygin gas playing the role of a classical scalar field, which in turn is linked to the $f(R)$ gravitational 
models through the correspondence described above, we reconstruct the $f(R)$ functionals in Section \ref{Originalmodel} for the 
original Chaplygin gas model and section \ref{Generalised} is devoted to $f(R)$ functionals for the generalized 
Chaplygin gas model. The general solutions are highly 
intractable, but we provide exact solutions in asymptotic regimes: early-time and far-future cosmologies. We have applied 
these solutions to de Sitter spacetimes, where exact potential forms as functions of both the scalar field 
and the scale factor are analysed, followed by discussions and conclusions in Section \ref{conc}.
%%%%%%%%%%%%%%%%%%%%%%%%%%%%%%%%%%%%%%%%%%%%%%%%%%%%%%%%%%%%%%%%%%%%%%%%%%%%%%%%%%%%%%%%%%%%%%%%%%%

In this paper, we will frequently use the natural units convention ($\hbar=c=k_{B}=8\pi G=1$)
and Latin indices $a\; ,b\; ,c\; ,\dots$ run from 0 to 3.
The symbols $\nabla$, $\D$ and the overdot $^{.}$ represent, respectively, the usual covariant derivative, the spatial covariant derivative, and differentiation with respect to cosmic time. We use the
$(-+++)$ spacetime signature and the Riemann tensor is defined by
\begin{eqnarray}
R^{a}_{bcd}=\Gamma^a_{bd,c}-\Gamma^a_{bc,d}+ \Gamma^e_{bd}\Gamma^a_{ce}-
\Gamma^f_{bc}\Gamma^a_{df}\; ,
\end{eqnarray}
where the $\Gamma^a_{bd}$ are the Christoffel symbols (i.e., symmetric in
the lower indices) defined by
\begin{equation}
\Gamma^a_{bd}=\frac{1}{2}g^{ae}
\left(g_{be,d}+g_{ed,b}-g_{bd,e}\right)\; .
\end{equation}
The Ricci tensor is obtained by contracting the {\em first} and the
{\em third} indices of the Riemann tensor:
\begin{equation}\label{Ricci}
R_{ab}=g^{cd}R_{cadb}\; .
\end{equation}
Unless otherwise stated, primes $^{'}\; ,^{''}$ etc are shorthands for derivatives with respect to the Ricci scalar
\begin{equation}
 R=R^{a}{}_{a}\;
\end{equation}
and $f$ is used as a shorthand for $f(R)$.
\section{Three alternative cosmological theories}\label{fRChapST}
In this section, we are going to briefly review the three alternative theories of gravity, namely $f(R)$ gravity, ST theory
and the Chaplygin-gas model. These gravity theories have been suggested and studied with the intention of showing that they can mimic both the
dark matter and dark energy mysteries. 
\subsection{Equivalence between $f(R)$ and scalar-tensor theories}
The scalar-tensor theory of gravity is a theory which tries to explain the interactions of gravity with matter. It is through this that the $f(R)$ gravity theory has been shown to be a sub-class of the ST theory \cite{faulk07}.
A clear illustration of how $f(R)$ theories are classified as a subclass of ST theory is in the BD theory for 
the case of the coupling constant $\omega=0$ \cite{sotiriou10}. 
The action in BD theory is given such that $\omega$ is independent of the scalar field \cite{clifton2011modified,sot10}.
Thus, we have
\begin{equation}
I_{BD}=\frac{1}{2\kappa}\int d^{4}x \sqrt{-g}\left[\phi R-\frac{\omega}{\phi}\nabla_{\mu}\phi\nabla^{\mu}\phi+\mathcal{L}_{m}(\Psi,g_{\mu\nu})\right]\; ,\label{stt2}
\end{equation} 
where $\kappa= 8\pi G$, $R$ is the Ricci scalar and $\mathcal{L}_{m}$ is the matter Lagrangian.
We consider the action that represents $f(R)$ gravity given as
\begin{equation}
I=\frac{1}{2\kappa}\int d^{4}x\sqrt{-g}\left[f(R)+\mathcal{L}_{m}\right]\; .
\end{equation}
The action in ST theory has the form \cite{amare1,scalar5}:
\begin{equation}
I_{f(\phi)}=\frac{1}{2\kappa}\int d^{4}\sqrt{-g}\left[f(\phi (R))+\mathcal{L}_{m}\right]\; ,\label{frstt1}
\end{equation}
where $f(\phi(R))$ is a function of $\phi(R)$ and we consider the scalar field $\phi$ to be \cite{scalar5,jos17}
\begin{equation}\label{fphi}
\phi=f'-1\; .
\end{equation}
Here, the scalar field $\phi$ should be invertible \cite{clifton12,sotiriou10,faulk07}.
Thus, if we compare the ST theory action to Eq. \eqref{stt2} of the BD theory  for the case of a vanishing coupling constant $\omega=0$, we can say 
that $f(R)$ theory is a special case of the ST theory. One can use the Palatini approach to show that $f(R)$ is a sub-class of ST theory but in that context, the coupling 
constant is considered to be $\omega=-\frac{3}{2}$ (see more detail in Refs. \cite{clifton12,scalar4}).

In this paper, we consider the treatment of the Chaplygin gas as a scalar fluid. With this in mind, we can make a basic analysis
of the Chaplygin gas through the ST theory of gravity. From the motivation that an equivalence between scalar-tensor
theory (BD-theory) and $f(R)$ theory of gravity exists, we obtain the energy density of the Chaplygin gas in terms of the scalar field, and from the Chaplygin gas property,
we obtain the Chaplygin gas pressure. The reason behind this treatment is that, in the literature, a lot of work has been done that suggests
that the Chaplygin gas can be treated as dark matter \cite{gorini04,dev03,kamenshchik2001alternative,bilic02,gorini03,ben12}.
\subsection{Chaplygin gas model}
In the original treatment, the negative pressure associated with the Chaplygin-gas
model is related to the (positive) energy density through the EoS \cite{elmardi16,gorini03,gorini04}
\begin{equation}\label{eq1}
p=-\frac{A}{\mu}\; ,
\end{equation}
where $A$ is a positive constant.
Using this equation of state in the conservation equation of $\mu$ given by
\begin{equation}\label{eq2}
\dot{\mu}+3\frac{\dot{a}}{a}\left(\mu+p\right)=0\; ,
\end{equation}
we find that the energy density of the Chaplygin gas evolves w.r.t the scale factor $a(t)$ as 
\begin{equation}\label{eq9}
\mu(a)= \sqrt{A+\dfrac{B}{a^{6}}}\; ,
\end{equation}
where $B$ is a constant of integration. One can see that for large $a(t)$, the energy density becomes independent of the scale factor for positive $B$.
This can be thought of as an empty universe with a cosmological constant (see more detailed analysis in Ref. \cite{gorini03}).
In Ref. \cite{gorini03}, it has been pointed out that for the early universe, the approximated expression of energy density can represent
a universe that contains pressureless dust matter.
From the Friedmann equation
\begin{equation}\label{eq10}
\Big(\dfrac{\dot{a}}{a}\Big)^2= \dfrac{\mu}{3}-\dfrac{k}{a^{2}}\; ,
\end{equation}
where $k$ stands for curvature and can be either $-1,0$ or $1$, depending on the spacetime geometry.
If we restrict ourselves to the case where the universe is flat, i.e., $k=0$, then
\begin{equation}\label{eq12}
\dot{a}=a\sqrt{\dfrac{\mu}{3}}\; .
\end{equation}
The energy density of the scalar field and pressure are given as \cite{ellis1991exact,gorini04}
\ber\label{eq13}
&&\mu_{\phi}= \dfrac{1}{2}\dot{\phi^{2}}+V(\phi)\;,\\
&&\label{eq14} p_{\phi}=\dfrac{1}{2}\dot{\phi^{2}}-V(\phi)\;.
\eer
Adding Eqs. \eqref{eq13} and \eqref{eq14} and using Eq. \eqref{eq9}, we get
\begin{equation}\label{eq17}
 \dot{\phi^{2}}= \frac{B}{a^3\sqrt{Aa^6+B}}\;.
\end{equation}
Using the following simple trick, where in this case, a prime denotes partial differentiation w.r.t the scale factor $a$,
\begin{equation}\label{eq18}
 \dot{\phi}= \dfrac{d\phi}{dt}= \dfrac{\partial \phi}{\partial a} \dfrac{\partial a}{\partial t}= \phi^{\prime} \dot{a}\;,
\end{equation}
and substituting the expression of \eqref{eq12}, we obtain
\begin{equation}\label{eq19}
\dot{\phi}= \phi^{\prime} a\sqrt{\dfrac{\mu}{3}}\;.
\end{equation}
Eq. \eqref{eq17} can therefore be rewritten as
\begin{equation}\label{eq21}
\phi^{\prime2}= \dfrac{3B}{a^{2}\Big(Aa^{6}+B\Big)}\implies \phi^{\prime}= \pm\dfrac{\sqrt{3B}}{a\sqrt{Aa^{6}+B}}\; .
\end{equation}
This equation can be integrated with respect to $a$ to yield
\begin{equation}\label{eq24}
\phi(a)=\pm\frac{1}{2\sqrt{3}}\ln\Big(\frac{Aa^{6}}{Aa^{6}+2B+2\sqrt{ABa^{6}+B^{2}}}\Big)+ C_{1}\; , 
\end{equation}
where $C_{1}$ is a constant of integration.
Our next step will be to reconstruct the $f(R)$ functional which produces $\phi$ per Eq. \eqref{fphi}. To do so, we first relate the Ricci scalar $R$ and the scale factor $a$ using the trace equation
\begin{equation}\label{eq30}
 R=\mu-3p=\frac{4A+\frac{B}{a^{6}}}{\sqrt{A+\frac{B}{a^{6}}}}\; ,
\end{equation}
which can also be rearranged as
\begin{equation}\label{eq33}
R a^{6}\sqrt{A\left(1+\frac{B}{Aa^{6}}\right)}-4Aa^{6}-B=0\; .
\end{equation}
From the equation describing energy density $\mu_{\phi}$ and pressure $p_{\phi}$, we solve for the potential as
\begin{equation}\label{eq34}
V(\phi)=\frac{\mu_{\phi}-p_{\phi}}{2}=\frac{2Aa^6+B}{2a^3\sqrt{Aa^6+B}}\; . 
\end{equation}
The de Sitter universe is a solution of the Einstein field equations with no standard matter sources but a positive cosmological constant or a scalar field. The Friedmann equation for such a universe reduces to
\begin{equation}\label{eq36}
 \Big(\frac{\dot a}{a}\Big) ^2 \equiv H^2 = \frac{\Lambda}{3}\; ,
\end{equation}
with the scale factor exponentially evolving as \cite{ellis1991exact,mendez1996exact}
\begin{equation}\label{eq37}
 a(t)= De^{\sqrt{\frac{\Lambda}{3}}t}=D e^{mt}\; .
\end{equation}
Using this solution for the scale factor at the early stages of the universe's evolution, 
one can obtain the characteristic potential that is solely dependent on the universe's cosmic time.
\section{Reconstruction of $f(R)$ gravity from the original Chaplygin gas model}\label{Originalmodel}
\subsection{Case 1: Early universe}
For the early universe, we assume that the scale factor $a$ is small enough for the approximation $Aa^6/B<<1$ to hold, that is, we can make the following treatment
\begin{equation}\label{eq38}
\sqrt{B^{2}\Big(1+\frac{Aa^{6}}{B}\Big)}\approx B\; . 
\end{equation}
Replacing this in Eq. \eqref{eq24}, we have
\begin{equation}\label{eq40}
\phi(a)= \mp \frac{1}{2\sqrt{3}}\ln\Big(1+\frac{4B}{Aa^{6}}\Big)+ C_{1}\; .
\end{equation}
From Eq. \eqref{eq33}, we have
\begin{equation}\label{eq41}
Ra^{3}\sqrt{B}-B=0\;.
\end{equation}
This means that we can write
\begin{equation}\label{eq44}
 a^{3}=\frac{\sqrt{B}}{R}\;,
\end{equation}
and hence replacing Eq. \eqref{eq44} in Eq. \eqref{eq40}, we have
\begin{equation}\label{eq46}
\phi(R)= \pm\frac{1}{2\sqrt{3}}\ln\Big(1+\frac{4R^{2}}{A}\Big)+ C_{1}\; .
\end{equation}
By substituting \eqref{eq46} into \eqref{fphi} and integrating, we obtain
\begin{equation}\label{eq49}
\begin{split}
 f(R)&=\pm \left[\frac{\sqrt{3}R}{6}\ln\Big(\frac{4R^{2} + A}{A}\Big) + \frac{\sqrt{3A}}{6}\arctan\Big(\frac{2R}{4A}\Big) 
 - \frac{\sqrt{3}R}{3} + C_{1}R\right] + R+C_{2}\; ,
\end{split}
\end{equation}
where $C_{2}$ is a constant of  integration.
\begin{figure}[h!]
\centering
\includegraphics[scale=0.6]{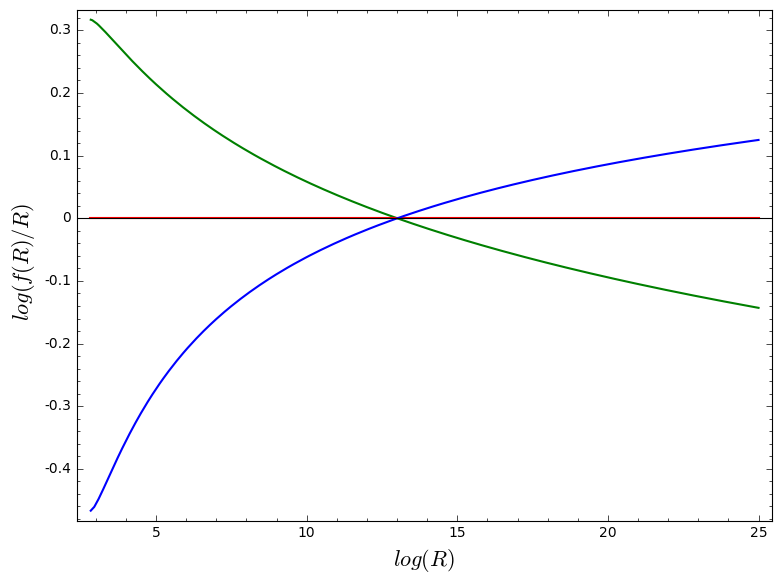}
\caption{Plot of $f(R)/R$ against Ricci scalar $R$ for the early universe, setting $A=0.5$, $C_{2}=0$ and $C_{1}=-0.4$, the red is for GR case, 
green is for positive $f(R)$ solution
and blue is for negative solution}
\label{fR1}
\end{figure}

Combining Eqs. \eqref{eq13} and \eqref{eq14} and use the fact that
\begin{equation}\label{eq53}
\sqrt{A+\frac{B}{a^{6}}}=\frac{1}{a^{3}}\sqrt{B}\sqrt{1+\frac{Aa^{6}}{B}}\approx  \frac{\sqrt{B}}{a^{3}}\; , 
\end{equation}
we obtain the potential $V(a)$ as
\begin{equation}\label{eq54}
V(a)=\frac{\sqrt{B}}{a^{3}}\; . 
\end{equation}
We get $a(\phi)$ from Eq. \eqref{eq40} as
\begin{equation}\label{eq55}
a^{3}=\frac{2\sqrt{B}}{\sqrt{A}(e^{\pm 2\sqrt{3}(\phi-C_{1})}-1)^{1/2}}\; . 
\end{equation}
Therefore, the potential is given as
\begin{equation}\label{eq56}
V(\phi)=\frac{\sqrt{A}}{2}\Big(e^{\pm 2\sqrt{3}(\phi-C_{1})}-1\Big)^{\frac{1}{2}}\; . 
\end{equation}
We set the constants such that $V(\phi)$ is given as
\begin{equation}\label{eq57}
V(\phi)=\frac{1}{2}\Big(e^{\pm 2\sqrt{3}\phi}-1\Big)^{\frac{1}{2}}\; . 
\end{equation}
Then the time-dependent potential can be obtained as
\begin{equation}\label{eq59}
 V(t) = \frac{\sqrt{B}}{D^{3}e^{3mt}} \; .
\end{equation}
By considering positive potential from Eq. \eqref{eq57}, one can write
\begin{equation}\label{positiveV(phi)}
V(\phi)=\frac{\sqrt{A}}{2}\Big(e^{2\sqrt{3}\phi}-1\Big)^{\frac{1}{2}}\; . 
\end{equation}
\begin{figure}[h!]
  \centering
  \subfloat[The potential $V(t)$ as a function of time $t$ from Eq. \eqref{eq59}
for values $A = B = m = D =1$ and $C = 0.$]{\includegraphics[width=0.5\textwidth]{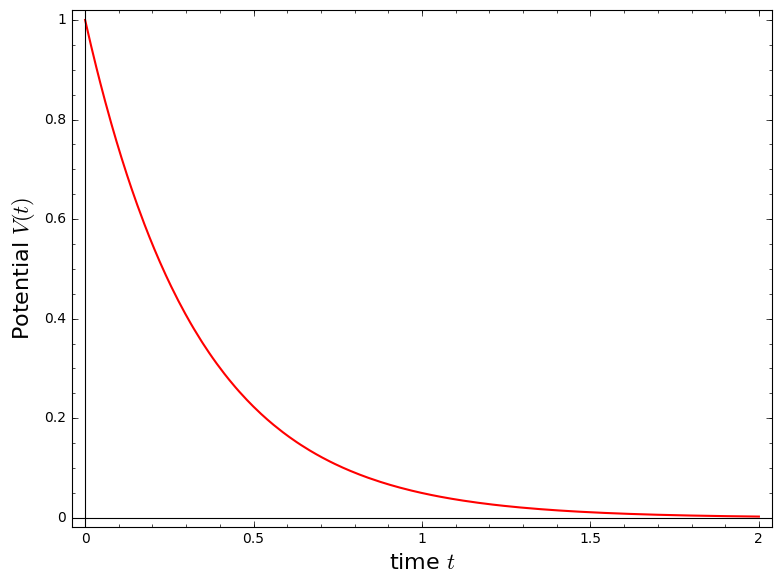}\label{fig:f1}}
 % \hfill
  \subfloat[The potential $V(\phi)$ as a function of the scalar field $\phi$ from Eq. \eqref{positiveV(phi)}
for values $A = B = m = D =1$ and $C = 0.$]{\includegraphics[width=0.5\textwidth]{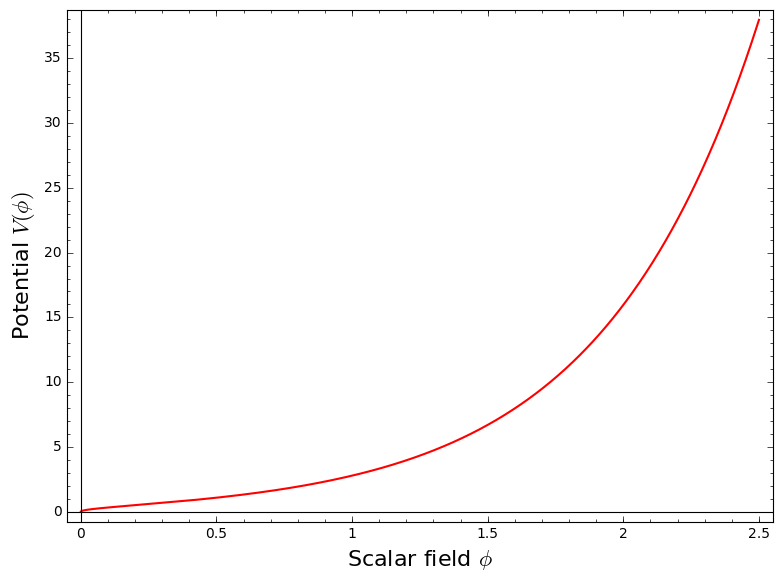}\label{fig:f2}}
  \caption{}
\end{figure}
\subsection{Case 2: Late universe}
The term on the denominator  \eqref{eq24} can be approximated with the assumption that for the late universe, the scale factor $a$ is large enough such that $B/Aa^{6}<<1$ holds. Therefore, we have the term as
\begin{equation}\label{eq60}
2\sqrt{ABa^{6}+B^{2}}=2\sqrt{ABa^{6}\Big(1+\frac{B}{Aa^{6}}\Big)} \approx 2\sqrt{AB}a^{3}\; .
\end{equation}
With this approximation, equation \eqref{eq24} becomes
\begin{equation}\label{eq63}
\phi(a)= \pm\frac{1}{2\sqrt{3}}\ln\Big(1+\frac{2B}{Aa^{6}}+\frac{2\sqrt{B}}{\sqrt{A}a^{3}}\Big)+ C_{1}\; . 
\end{equation}
Therefore Eq. \eqref{eq63} takes the form
\begin{equation}\label{eq64}
\phi(a)= \pm \frac{1}{2\sqrt{3}}\ln\Big(\frac{2\sqrt{B}}{\sqrt{A}a^{3}}\Big)+ C_{1}\;,
\end{equation}
 Eq. \eqref{eq33} results in
\begin{equation}\label{eq66}
a^{6}=\frac{\sqrt{B}}{\sqrt{R\sqrt{A}-4A}}\; ,~~R\neq 4\sqrt{A}. 
\end{equation}
Therefore, we update Eq. \eqref{eq63} as
\begin{equation}\label{63new}
\phi(R)=\pm \frac{1}{2\sqrt{3}}\ln \Big(1+\frac{2}{\sqrt{A}}\sqrt{R\sqrt{A}-4A}\Big)+C_{1}\; . 
\end{equation}
Replacing Eq. \eqref{63new} in Eq. \eqref{fphi}, we have $f(R)$ as
\begin{equation}\label{eq77}
\begin{split}
f(R)&=\pm\Big[\frac{\sqrt{3}}{6}\Big(R - \frac{17\sqrt{A}}{4}\Big)\ln\Big(1+ \frac{2\sqrt{R\sqrt{A} - 4A}}{\sqrt{A}}\Big) 
+ \frac{\sqrt{3}}{12}\sqrt{R\sqrt{A} - 4A} + \Big(-\frac{\sqrt{3}}{12} + C_{1}\Big)R \\
&+ \frac{19\sqrt{3A}}{48}\Big] + R +C_{3}\; , 
\end{split}
\end{equation}
where $C_{3}$ is a constant of integration.
\begin{figure}[h!]
\centering
\includegraphics[scale=0.6]{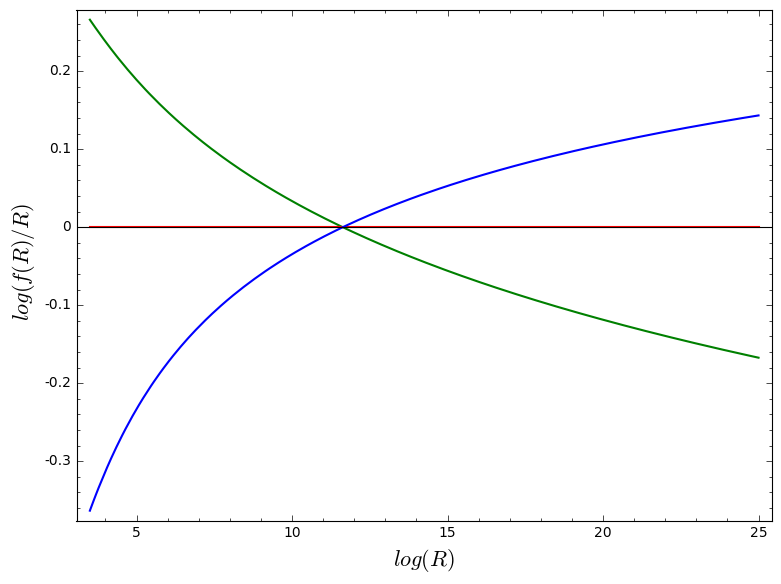}
\caption{Plot of $f(R)/R$ against Ricci scalar $R$ for the late universe, setting $A=0.4$,  $C_{3}=0$ and $C_{1}=-0.4$, the red line represents the GR case, while the green and blue curves depict negative and positive $f(R)$ solutions respectively}
\label{fR2}
\end{figure}
By combining Eqs. \eqref{eq13} and \eqref{eq14}  and using the large-scale-factor approximation,
we have the potential as
\begin{equation}\label{eq81}
V(\phi)=\sqrt{A}\; .
\end{equation}
As a combination of Eqs. \eqref{eq13} and \eqref{eq14} results in
\begin{equation}\label{eq78}
 V(a)=\frac{2A+\frac{B}{a^{6}}}{2\sqrt{A+\frac{B}{a^{6}}}}\; ,
\end{equation}
one can start from dependence of scale factor $a$ on scalar field $\phi$ to have the potential $V(\phi)$ as
\begin{equation}\label{newV(phi)}
V(\phi)=\frac{2A+\frac{A}{4}\Big(e^{\mp 2\sqrt{3}\phi-C_{1}}\Big)^{2}}{2\sqrt{A+\frac{A}{4}\Big(e^{\mp 2\sqrt{3}\phi-C_{1}}\Big)^{2}}}\; . 
\end{equation}
In principle, we can constrain the scalar field $\phi$ by equating Eqs. \eqref{eq81} and \eqref{newV(phi)}. Once this is done, one has a constant scalar field given by
\begin{equation}
\phi=\pm \frac{C_{1}}{2\sqrt{3}}\; . 
\end{equation}
So far, we have considered the original model of the Chaplygin gas and have obtained expressions for $f(R)$ that correspond to the early and late
universe along with their corresponding potentials. In the following, we present a similar analysis
for the generalized Chaplygin model.

\section{Reconstruction of $f(R)$ gravity from the generalized Chaplygin gas model}\label{Generalised}
For the generalized Chaplygin gas, we have  the pressure as \cite{elmardi16,gorini03,gorini04}
\begin{equation}\label{eq85}
 p=-\frac{A}{\mu^{\alpha}}\; ,
\end{equation}
where $0\leq \alpha \leq 1$ and $A$ is a constant. 
% The energy density as a function of the scale factor for this model is given by \cite{gorini03,gorini04}
\begin{equation}\label{eq88}
\mu(a)=\Big(A + Ba^{-3(\alpha+1)}\Big)^{\frac{1}{1+\alpha}}\; . 
\end{equation}
For a flat universe, $K=0$, we can write the Friedmann equation as
\begin{equation}\label{eq89}
\dot{a}=a\sqrt{\frac{\mu}{3}}\; . 
\end{equation}
Combining Eqs. \eqref{eq13} and \eqref{eq14} results in
\begin{equation}\label{eq91}
\dot{\phi}^{2}=\frac{Ba^{-3(\alpha+1)}}{\Big(A+Ba^{-3(\alpha+1)}\Big)^{\frac{\alpha}{1+\alpha}}}\; . 
\end{equation}
Therefore, we can write
\begin{equation}\label{eq93}
\phi'=\pm\sqrt{\frac{3B}{Aa^{3\alpha + 5}+Ba^{2}}}\; . 
\end{equation}

\subsection{Case1: Early universe}
We consider Eq. \eqref{eq93}, for the early universe, the scale factor is assumed to be small enough, therefore,
we can write 
\begin{equation}
d\phi = \pm\frac{\sqrt{3B}}{\sqrt{Ba^{2}(1+\frac{Aa^{3(\alpha+1)}}{B})}}da\; , 
\end{equation}
then using the fact that $(1+x)^{n}\approx 1+nx +O(x^{2}), \text{ for } x<<1$, we have
\begin{equation}\label{eq110}
d\phi = \pm\frac{\sqrt{3}}{a(1+\frac{Aa^{3(\alpha+1)}}{2B})}da\; . 
\end{equation}
Integrating this equation, we have
\begin{equation}\label{eq111}
\phi(a)=\pm\frac{\sqrt{3}}{3(\alpha+1)}\ln \Big(\frac{2B}{a^{3(\alpha+1)}}+A\Big)+C_{1}\; . 
\end{equation}
By imposing the early universe assumption for scale factor on Eq. \eqref{eq96}, we have $a(R)$ as
\begin{equation}\label{eq112}
a(R)=\frac{B^{\frac{1}{3(\alpha+1)}}}{R^{1/3}}\; . 
\end{equation}
Therefore, we have $\phi(R)$ as 
\begin{equation}\label{eq113}
\phi(R)=\pm\frac{\sqrt{3}}{3(\alpha+1)}\ln (2R^{\alpha+1}+A)+C_{1}\; . 
\end{equation}
Thus integrating with respect to the Ricci scalar $R$ from
\begin{equation}\label{eq114}
f(R)=\int \phi dR+R+C_{4}\; , 
\end{equation}
to have
\begin{equation}\label{eq115}
f(R)=\pm \left[\frac{\sqrt{3} \Big(R(\alpha + 1) -R \ln \left(2R^{\alpha+1}+A\right)\Big)}{3(\alpha+1)}
-\frac{\sqrt{3}A}{3}
\int \frac{dR}{2R^{\alpha+1} + A}+ C_{1}R\right] + R +C_{4}\; , 
\end{equation}
where $C_{4}$ is a constant the integration.
\begin{figure}[h!]
\centering
\includegraphics[scale=0.6]{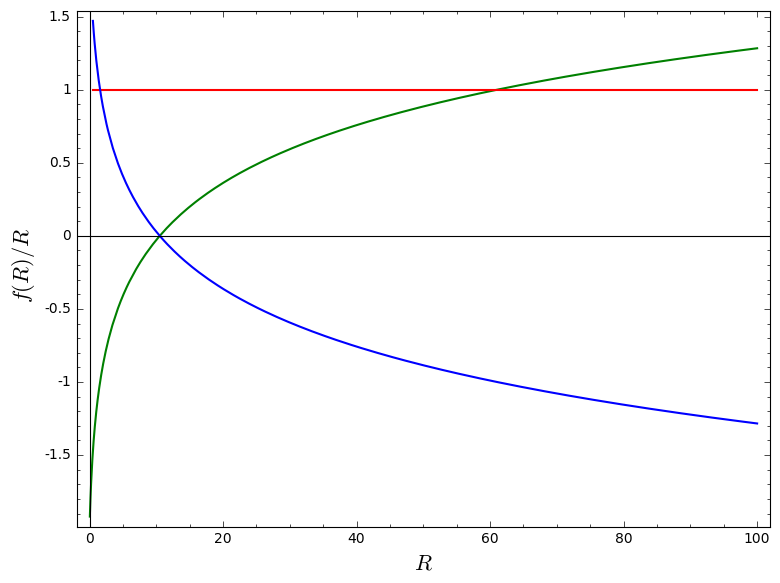}
\caption{Plot of $f(R)/R$ against Ricci scalar $R$ for the early universe, setting $A=0$, $C_{4}=0$, $C_{1}=0.4$ and $\alpha=0.2$ the red line represents the GR case, while the green and blue curves depict negative and positive $f(R)$ solutions respectively}
\label{fR3}
\end{figure}
The potential is given by
\begin{equation}\label{eq116}
V(\phi)=\frac{\mu_{\phi}-p_{\phi}}{2}\; . 
\end{equation}
Considering energy density as
\begin{equation}\label{eq117}
\mu=\Big(A+Ba^{-3(\alpha+1)}\Big)^{\frac{1}{\alpha+1}} 
\end{equation}
and pressure as
\begin{equation}\label{eq118}
 p=-\frac{A}{\mu^{\alpha}}\; ,
\end{equation}
we have
\begin{equation}\label{eq119}
V(a)=\frac{2A+\frac{B}{a^{3(\alpha+1)}}}{2(A+Ba^{-3(\alpha+1)})^{\alpha/(\alpha+1)}}\; . 
\end{equation}
By treating the denominator, we have
\begin{equation}\label{eq120}
(A+Ba^{-3(\alpha+1)})^{\frac{\alpha}{(\alpha+1)}}=\Big(\frac{B}{a^{3(\alpha+1)}}(\frac{a^{3(\alpha+1)}}{B}+1)\Big)^{\frac{\alpha}{\alpha+1}}
\approx \Big(\frac{B}{a^{3(\alpha+1)}}\Big)^{\frac{\alpha}{\alpha+1}}\; ,
\end{equation}
where we have considered that $a^{3(\alpha+1)}$ is small enough in the early universe.
So the potential becomes
\begin{equation}\label{eq121}
V(a)=\frac{2A+\frac{B}{a^{3(\alpha+1)}}}{2\Big(\frac{B}{a^{3(\alpha+1)}}\Big)^{\frac{\alpha}{\alpha+1}}}\; . 
\end{equation}
From Eq. \eqref{eq111}, we have
\begin{equation}\label{eq123}
a^{3(\alpha+1)}=\frac{2B}{e^{\mp\frac{3\phi(\alpha+1)}{\sqrt{3}}}-A}\; .
\end{equation}
If we substitute  \eqref{eq123} into \eqref{eq121}, we get the potential $V$ as a function of $\phi$ as 
\begin{equation}\label{eq124a}
V(\phi)= \frac{1}{2^{\frac{\alpha+2}{\alpha+1}}B}\Big(2A+\frac{1}{2}(e^{\pm\frac{3\phi(\alpha+1)}{\sqrt{3}}}-A)\Big)
\Big(e^{\pm\frac{3\phi(\alpha+1)}{\sqrt{3}}}-A\Big)^{\frac{1}{\alpha+1}}\; .
\end{equation}
We only consider positive potential as
\begin{equation}\label{eq124}
V(\phi)= \frac{1}{2^{\frac{\alpha+2}{\alpha+1}}B}\Big(2A+\frac{1}{2}(e^{\frac{3\phi(\alpha+1)}{\sqrt{3}}}-A)\Big)
\Big(e^{\frac{3\phi(\alpha+1)}{\sqrt{3}}}-A\Big)^{\frac{1}{\alpha+1}}\; .
\end{equation}
Subsequently, Equations \eqref{eq121} and \eqref{eq37} give rise to
\begin{equation}\label{eq126}
V(t)=\frac{1}{2B^{\frac{\alpha}{\alpha+1}}}\Big(2A+B(De^{mt})^{-3(\alpha+1)}\Big)(De^{mt})^{3\alpha}\; . 
\end{equation}

\begin{figure}[h!]
  \centering
    \subfloat[The potential $V(\phi)$ as a function of scalar field $\phi$ from Eq. \eqref{eq124}
for values $A = B = m = D =1$ and $\alpha=0.5$ (red), $\alpha=0.3$ (blue) and $\alpha=0.2$ (green).]
 {\includegraphics[width=0.5\textwidth]{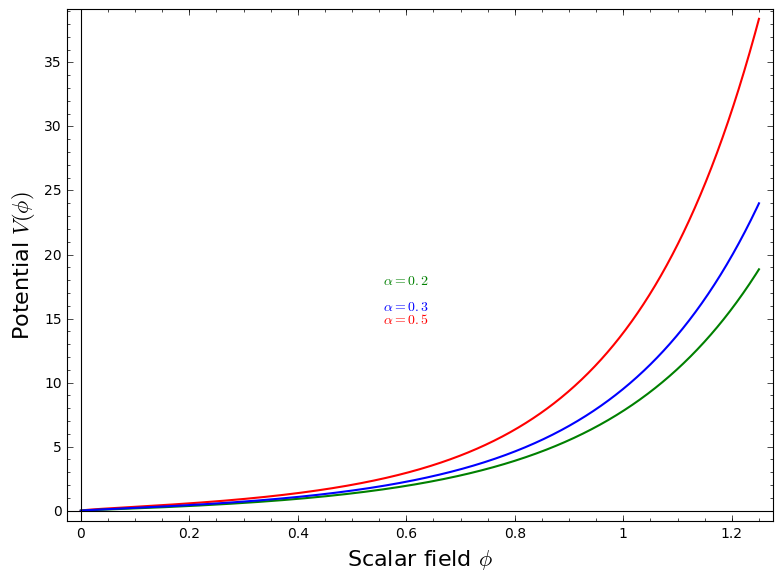}\label{fig:a}}
 % \hfill
  \subfloat[The potential $V(t)$ as a function of time $t$ from Eq. \eqref{eq126}
for values $A = B = m = D =1$ and $\alpha=0.5$ (red), $\alpha=0.3$ (blue) and $\alpha=0.2$ (green).]{\includegraphics[width=0.5\textwidth]{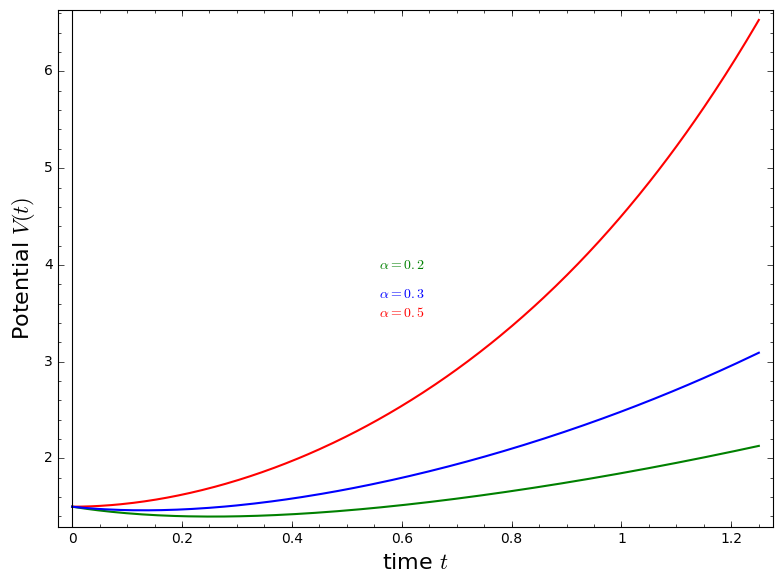}\label{fig:b}}
  \caption{}
\end{figure}

\subsection{Case2: Late universe}
For the late universe, we have the following assumption
\begin{equation}\label{eq95}
Ba^{2}<< Aa^{3\alpha+5}\; . 
\end{equation}
Thus we have
\begin{equation}\label{eq96}
d\phi \approx  \sqrt{\frac{3B}{Aa^{3\alpha + 5}}}da\; .
\end{equation}
We perform integration to get
\begin{equation}\label{eq97}
\phi(a)=\pm \frac{2\sqrt{B}}{\sqrt{3}\sqrt{A}(\alpha+1)a^{3(\alpha+1)/2}}+C_{1}\;.
\end{equation}
The next step is to get $\phi(R)$. Using the trace equation once again yields
\begin{equation}\label{eq99}
R=\frac{Ba^{-3(\alpha+1)}+4A}{\Big(A + Ba^{-3(\alpha+1)}\Big)^{\frac{\alpha}{1+\alpha}}}\; . 
\end{equation}
By manipulating Eq. \eqref{eq99}, we obtain
\begin{equation}\label{eq100}
a(R)=\Big(\frac{RA^{\frac{\alpha}{1+\alpha}}-4A}{B(1-\frac{\alpha R}{1+\alpha}A^{\frac{-1}{\alpha+1}})}\Big)^{-\frac{1}{3(\alpha+1)}}\; .
\end{equation}
Thus from Eq. \eqref{eq97}, we have
\begin{equation}\label{eq101}
\phi(R)=\pm\frac{2\sqrt{B}}{\sqrt{3A}(\alpha+1)}\Big(\frac{RA^{\frac{\alpha}{1+\alpha}}-4A}{B(1-\frac{\alpha R}{1
+\alpha}A^{\frac{-1}{\alpha+1}})}\Big)^{\frac{1}{2}}\; . 
\end{equation}
Substituting this expression of $\phi(R)$ in Eq. \eqref{fphi}, we get 
\begin{equation}\label{eq103}
\begin{split}
f(R)=&\pm\Bigg[\frac{-A^{2}(3\alpha-1)^{2}m_{1}(A,\alpha,R)\arcsin\Big(\frac{2\alpha A\frac{\alpha}{\alpha+1}R-5A\alpha
-A}{A(3\alpha-1)}\Big)+m_{2}(A,\alpha,R)}{4\sqrt{3}\alpha^{3/2}(\alpha+1)^{3/2}m_{1}(A,\alpha,R)} +C_{1}R\Bigg]\\
&+ R +C_{5}\; , 
\end{split}
\end{equation}
where $C_{5}$ is a constant of integration ,
\begin{equation}
m_{1}(A,\alpha,R)=\sqrt{-\alpha A^{\frac{\alpha}{\alpha+1}} \big(\alpha R
-(\alpha+1)A^{\frac{1}{\alpha+1}}\big)\big(RA^{\frac{\alpha}{\alpha+1}}-4A\big)}\; , 
\end{equation}
and 
\begin{equation}
\begin{split}
 m_{2}(A,\alpha,R)=&-4\alpha^{3}A^{\frac{3\alpha}{\alpha+1}}R^{3}+6(5\alpha+1)\alpha^{2}A^{\frac{3\alpha+1}{\alpha+1}}R^{2}
-2\alpha (33\alpha^{2}+18\alpha+1)A^{\frac{3\alpha+2}{\alpha+1}}R\\
&+(40\alpha^{3}+48\alpha^{2}+18\alpha)A^{3}\; .
\end{split}
\end{equation}
\begin{figure}[h!]
\centering
\includegraphics[scale=0.6]{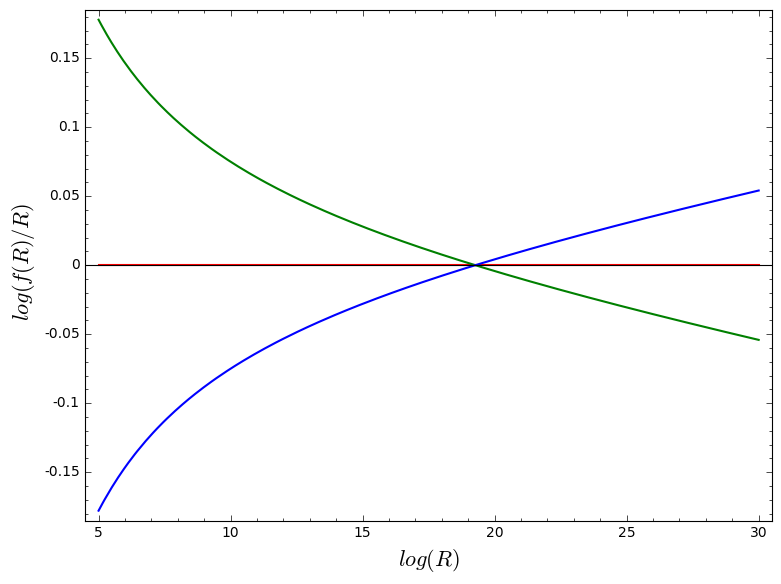}
\caption{Plot of $f(R)/R$ against Ricci scalar $R$ for the late universe, setting $A=10^{-4}$, $C_{1}=-2\times 10^{-3}$, $C_{5}=1$ 
and $\alpha=3\times 10^{-5}$ the red line represents the GR case, while the green and blue curves depict negative and positive $f(R)$ solutions respectively}
\label{fR4}
\end{figure}

From Eq. \eqref{eq88} in late universe, that is together with its corresponding pressure, we have potential $V(a)$ from Eq. \eqref{eq34} as
\begin{equation}\label{eq105}
V(a)=\frac{Ba^{-3(\alpha+1)}+A}{2\big(A+Ba^{-3(\alpha+1)}\big)^{\frac{\alpha}{\alpha+1}}}\; . 
\end{equation}
Replacing Eq. \eqref{eq97} in Eq. \eqref{eq105} we get
\begin{equation}\label{eq106}
V(\phi)=\frac{3A(\alpha+1)^{2}\phi^{2}+8AB}{8B(A+3A(\alpha+1)^{2}\phi^{2})^{\frac{\alpha}{\alpha+1}}}\; .
\end{equation}
We replace scale factor \eqref{eq37} in \eqref{eq105} to get potential as a function of time as
\begin{equation}\label{eq108}
V(t) = \frac{B(De^{mt})^{-3(\alpha+1)}+2A}{2\big(A+B(De^{mt})^{-3(\alpha+1)}\big)^{\frac{\alpha}{\alpha+1}}}\; .
\end{equation}

\begin{figure}[h!]
  \centering
  \subfloat[The potential $V(\phi)$ as a function of scalar field $\phi$ from Eq. \eqref{eq106}
for values $A = B = m = D =1$, and $\alpha=0.5$ (red), $\alpha=0.3$ (blue) and $\alpha=0.2$ (green).]{\includegraphics[width=0.5\textwidth]{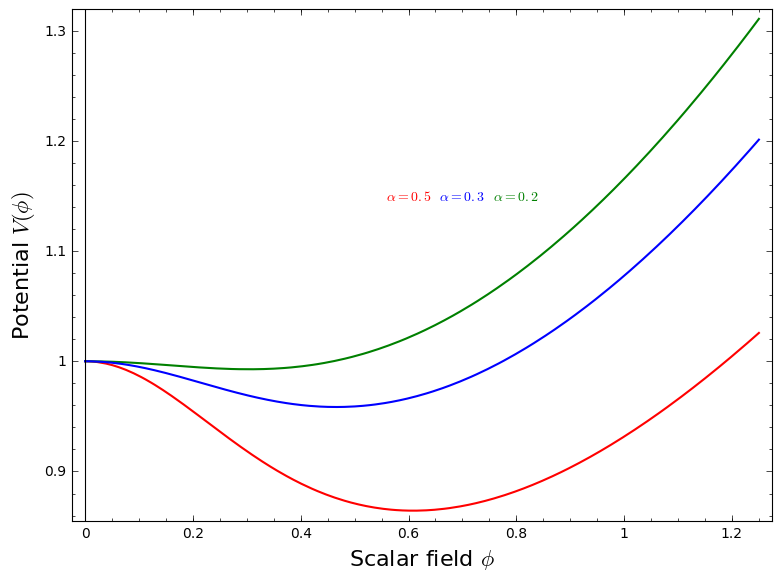}\label{fig:a}}
 % \hfill 
    \subfloat[The potential $V(t)$ as a function of time $t$ from Eq. \eqref{eq108}
for values $A = B = m = D =1$, and $\alpha=0.5$ (red), $\alpha=0.3$ (blue) and $\alpha=0.2$
 (green).]
 {\includegraphics[width=0.5\textwidth]{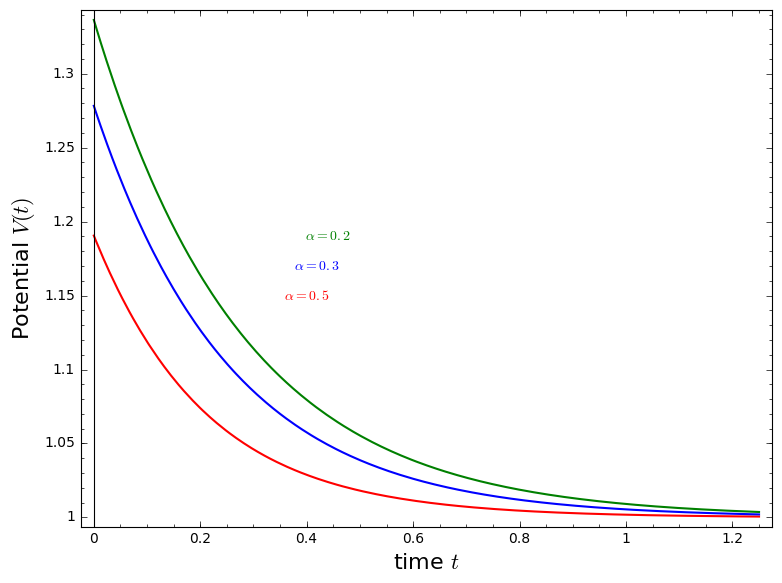}\label{fig:b}}
  \caption{}
\end{figure}
For the early universe, we plotted the potential dependence on the scalar field and one can easily observe that the potential increases with an increase in $\phi$.  However, for the late universe, this dependence is realized after a decrease to a certain minimum. It can be observed that as $\alpha$ increases the 
position of the minimum point gets lowered. For the time-dependent potential in the early universe,
an increase in time results in an increase in the potential. Conversely, the potential decreases as
time increases for the late universe. This trend has been obtained through the use of approximated solutions.
The actual behavior of the potential under consideration may be obtained through acquiring the exact solutions of the scalar field.

\section{Conclusions}\label{conc}
The reconstruction of $f(R)$ models from the Chaplygin gas cosmological model is systematically developed in this work. This was done through the consideration of the equivalence between two theories of gravitation, namely $f(R)$ theory and the Brans-Dicke subclass of scalar-tensor models. The analysis  made use of a combination of energy density, pressure, kinetic and potential terms of the scalar field to obtain relationships between
the scalar field, the Ricci scalar and the scale factor. As can be seen from the reconstructed functionals and their corresponding plots versus the GR Lagrangian density ($f(R)=R)$, these expressions  have GR as their respective  limiting solutions.

The results have been applied to a de Sitter universe and the time-dependent expressions for the 
potential have been obtained for both the original and generalized cases of the Chaplygin gas model. 
This was made possible by two asymptotic assumptions, early- and late-universe epochs of cosmic evolution. The behaviors of the potential for both the original and generalized cases show that the amplitude of the potential gets weaker with time in a de Sitter universe except for the early universe consideration in the generalized Chaplygin gas model where
the potential grows with time. However, the dependence of the potential on the scalar field shows that for larger values of
the scalar field, the potential is also large, but for the generalized case within a late universe, the potential is
observed to have a valley before it follows the trend observed for other potentials.  

The solutions provided in this work are obtained for asymptotic cases, but intermediate solutions are highly intractable at best. We thus leave the (numerical) analysis of solutions for the entire cosmic history for a future work.

\section*{Acknowledgements}
This work is based on the research supported in part by the National Research Foundation of South Africa (Grant Number 109257).
AA  also acknowledges the Faculty Research Committee of the Faculty of Agriculture, Science and Technology of North-West University
for financial support. JN gratefully acknowledges financial support from the Swedish International Development Cooperation Agency (SIDA) 
through the International Science Program (ISP) to the University of Rwanda (Rwanda Astrophysics, Space and Climate Science Research Group),
and Physics Department, North-West University, Mafikeng
Campus, South Africa, for hosting him during the preparation of this paper.

\bibliographystyle{unsrt}
\bibliography{bibliography}

\end{document}